# DefoEye: Python-Based Software for Facilitating Time-Series InSAR Analysis of Sentinel-1 Remote-Sensing Data

Alireza Taheri Dehkordi[1, 2, *], Hossein Hashemi[1, 2, 3], Amir Naghibi[1, 2, 3]

[1] Division of Water Resources Engineering (TVRL), Faculty of Engineering (LTH), Lund university, Lund, Sweden.

[2] United Nations University Hub on Water in a Changing Environment (WICE), United Nations University Institute for Water, Environment and Health (UNU-INWEH), Lund University, Lund, Sweden.

[2] Centre for Advanced Middle Eastern Studies, Lund university, Lund, Sweden.

[*] Corresponding author (alireza.taheri_dehkordi@tvrl.lth.se)

## Abstract

Time-Series Interferometric Synthetic Aperture Radar (TS-InSAR) is a powerful remote sensing technique for monitoring Ground Deformation (GD), which poses threats to geological stability and civil infrastructure worldwide. However, many existing TS-InSAR processing software tools face limitations, such as restricted geographic applicability, commercial licensing, and a lack of full end-to-end processing workflow support. While GMTSAR avoids some of these challenges, it still relies heavily on manual user intervention, requires the execution of C-shell commands, lacks a user-friendly GUI, and misses key processing steps such as interferogram network pruning and anchoring of unwrapped interferograms, all of which hinder both accessibility and accuracy. This paper introduces DefoEye (v1), an open-source Python-based software that wraps GMTSAR to facilitate the full TS-InSAR processing chain for Sentinel-1 data through a user-friendly interface. DefoEye provides an end-to-end workflow that includes parallel job execution across different steps, interferogram network pruning, and multiple anchoring options. Its applicability was evaluated in four regions with differing geological settings, causes of GD, atmospheric and climatic conditions over the 2020–2024 period. For three regions—Bologna (Italy), Gotland (Sweden), and Houston (USA)—DefoEye results were compared against 10 available GNSS stations, demonstrating strong consistency (RMSE: 4.3–11.9 mm; PCC: 0.63–0.95). For Karaj (Iran), where no GNSS stations were available, DefoEye results were compared against other widely used processing tools, yielding similarly close agreement (RMSE: 4.8 mm/yr; PCC: 0.98), with DefoEye conducting the process through a unified workflow. These findings show that DefoEye delivers reliable TS-InSAR results suitable for a variety of geological, hydrological, and environmental applications. DefoEye software and a detailed step-by-step tutorial are available at: https://github.com/ATDehkordi/DefoEye.

# 1. Introduction

Ground deformation (GD) is a critical geohazard caused by both natural processes, such as tectonic activity, and human activities, such as groundwater extraction (Smith et al., 2021, Bui et al., 2021). Continuous monitoring of GD is essential for understanding its spatial and temporal dynamics to assess potential hazards and enhance regional safety (Usha et al., 2025). Traditional geodetic methods, such as leveling and Global Navigation Satellite System (GNSS) stations, offer high accuracy but are limited by sparse spatial coverage (Li et al., 2024). In contrast, satellite-based Interferometric Synthetic Aperture Radar (InSAR) is a powerful Remote Sensing (RS) tool for GD mapping, enabling coverage of extensive geographical areas (Jaramillo et al., 2024).

InSAR utilizes phase information from Synthetic Aperture Radar (SAR) images to detect Earth's surface changes along the off-nadir direction of the satellite's Line of Sight (LOS) (Teixeira et al., 2024). Conventional InSAR, referred to as Differential InSAR (D-InSAR), relied on a limited number of images and was hindered by several challenges, such as atmospheric phase delays and spatiotemporal incoherence (Wang et al., 2020). To address these limitations, InSAR techniques have evolved from D-InSAR to Time-Series InSAR (TS-InSAR), which utilizes multitemporal SAR images and provides enhanced monitoring accuracy and continuity (Xiong et al., 2025).

Current TS-InSAR processing platforms often suffer from key limitations that hinder both accessibility and adoptability. For instance, commercial software such as GAMMA (Wegnüller et al., 2016) and SARPROZ (Perissin et al., 2011) require paid licenses. Some software, such as LiCSBAS, is limited in applicability to geographic regions where pre-generated LiCSAR products are available (Morishita et al., 2020). Moreover, some platforms do not support end-to-end processing workflows; for example, MintPy (Yunjun et al., 2019) depends on externally processed inputs, forcing users to combine multiple processing platforms—often resulting in format incompatibility and workflow inefficiencies (Zaki et al., 2024, Hrysiewicz et al., 2023). Additionally, some software, such as GIAnT, are outdated, poorly maintained, and incompatible with modern programming languages like Python 3.x, reducing their long-term usability (Agram et al., 2013). These challenges clearly indicate the need for a freely accessible, open-source, globally applicable, end-to-end TS-InSAR processing solution to meet the demands of the big SAR data era (Li et al., 2022, Zaki et al., 2024).

Unlike other InSAR processing tools mentioned above, the open-source Generic Mapping Tools SAR (GMTSAR) software includes an end-to-end TS-InSAR analysis workflow, covering all processing steps from co-registration to the final retrieval of the GD map (Sandwell et al., 2011b). It is also globally applicable and not geographically restricted. Hence, it addresses many of the above-mentioned challenges that other InSAR software face. These advantages make GMTSAR a suitable choice for TS-InSAR analysis. However, running GMTSAR requires extensive familiarity with C-shell scripting in the UNIX/LINUX operating systems, as all functions and parameters are controlled via the command line, which presents a significant barrier for the users (Sandwell et al., 2011a). Additionally, GMTSAR is highly user-driven, meaning that users must manually run various parts of the processing workflow, which is not only demanding and time-consuming but also challenging for those without prior experience in RS or programming. Moreover, GMTSAR lacks key processing steps in its workflow, such as interferogram network pruning and anchoring of unwrapped interferograms to a reference point, which is essential for meaningful deformation interpretation (Haghshenas Haghighi and Motagh, 2024a, Morishita et al., 2020).

To address the above-mentioned limitations, this paper presents DefoEye (v1), an open-source Python (3.x)-based software that wraps GMTSAR to provide enhanced and streamlined TS-InSAR workflow. The use of Python scripts offers high flexibility and repeatability, enabling continuous improvement and easy integration of future enhancements (Licata et al., 2025). The main contributions of DefoEye can be summarized as:

- A streamlined, end-to-end TS-InSAR workflow that wraps GMTSAR, from SAR data download to final GD map generation and export, implemented through a structured sequence of processing steps and a user-friendly Graphical User Interface (GUI).
- Parallelized job processing of key computational steps, including image co-registration, interferogram generation and correction, sub-swath merging, and phase unwrapping, significantly accelerating TS-InSAR analysis.
- Newly added features including automatic directory setup, interferogram network pruning using triplet loop closure, flexible anchoring methods (including point/region referencing and average-phase referencing), self-checks at different steps to ensure correct execution, and an enhanced export module with automatic validation using GNSS.

To assess its trans-applicability and generalization capability, DefoEye was tested at four study sites with differing geological settings, causes of GD, atmospheric and climatic conditions: Bologna (Italy), Gotland (Sweden), Houston (USA), and Karaj (Iran). The performance of DefoEye was validated by comparing its results with data from multiple GNSS stations and with the results of other TS-InSAR processing software. The remainder of this paper is structured as follows: Section 2 outlines the design and development of DefoEye. Section 3 describes the implementation of the software across three study areas. Section 4 provides a detailed discussion of the findings, followed by concluding remarks in Section 5.

# 2. Design and development

## 2.1. Data sources

With its improved orbital control, Sentinel-1 (S1) minimizes noise contributions and mitigates decorrelation issues associated with long spatial and temporal baselines in previous SAR missions (Tang et al., 2023). S1 data is freely available and publicly accessible, continuously monitoring Earth since 2014 with Sentinel-1A (S1A), followed by Sentinel-1B (S1B) in 2016 (failed in 2021). With the recent launch of Sentinel-1C (S1C) in late 2024, it has become one of the longest-operating SAR satellite missions. Given these advantages, DefoEye (v1) provides a facilitated workflow for TS-InSAR analysis of S1 data.

S1 mission is operated in a sun-synchronous orbit by the European Space Agency (ESA), acquiring SAR data in the C-band, with a wavelength of approximately 5.6 cm (Torres et al., 2012). Each of the S1 satellites (S1A, S1B, and S1C) provides a 12-day revisit capability. For InSAR applications, S1 Single Look Complex (SLC) products must be used, as they contain both amplitude and phase information acquired in Interferometric Wide (IW) mode (Morishita et al., 2020). The IW mode utilizes a burst radar acquisition technique called Terrain Observation by Progressive Scans (TOPS), consisting of three sub-swaths (IW1, IW2, and IW3), each composed of a series of bursts. The dataset has a spatial resolution of approximately 5 × 20 meters (azimuth × range directions), and a swath width of 250 km. S1 IW images are captured in dual polarization—vertical transmit and vertical receive (VV), and vertical transmit and horizontal receive (VH). However, for InSAR applications, VV polarization is used due to its lower susceptibility to temporal decorrelation (Tang et al., 2023). DefoEye includes an easy-to-use downloader module for the automatic retrieval of S1 data from the Alaska Satellite Facility (ASF) SAR Distributed Active Archive Center (DAAC) (https://asf.alaska.edu/asfsardaac/, last accessed March 2025). In addition, DefoEye automatically downloads S1 Precise Orbit Determination (POD) files from the Copernicus service (https://browser.dataspace.copernicus.eu).

The Shuttle Radar Topography Mission (SRTM) Global Digital Elevation Model (DEM), with a spatial resolution of 30 meters (one arc-second), is automatically downloaded within the processing chain of DefoEye to align S1 TS data to a reference image, flatten interferograms, and geocode the results (Farr et al., 2007). Further details are provided in Section 2.2.

To evaluate the performance of the proposed software, TS-InSAR results can be compared with an independent external data source. DefoEye includes an automated validation module based on GNSS data from the Nevada Geodetic Laboratory (NGL) (https://geodesy.unr.edu/index.php, last accessed March 2025), which offers global GNSS observations (Blewitt et al., 2018). The proposed validation module is compatible with daily GNSS solutions in the tenv3 format, provided in the International GPS Service Reference Frame 2014 (IGS14).

## 2.2. Processing workflow

DofoEye and a step-by-step tutorial illustrating the different stages of the software can be found here: https://github.com/ATDehkordi/DefoEye. The proposed software provides a user-friendly GUI to enhance user interaction. In the following, the TS-InSAR processing chain of DefoEye (Fig. 1) is fully explained.

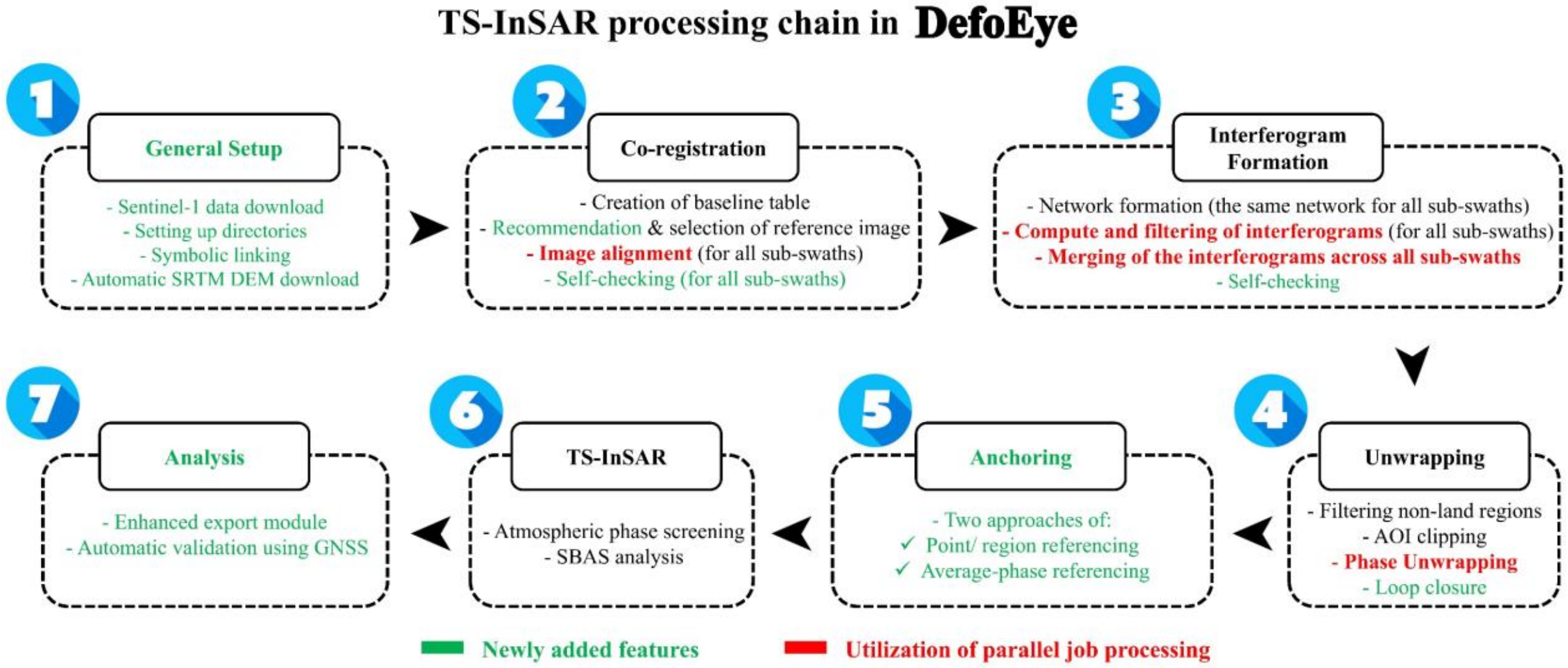

**Figure 1.** TS-InSAR processing chain of DefoEye (v1). Green text represents newly added features in DefoEye for wrapping GMTSAR, while red text indicates steps where parallel job processing is utilized.

### *2.2.1. General Setup*

DefoEye provides an easy-to-use module for the automatic download of S1 data and the corresponding POD files. Alternatively, if these files have already been obtained, they user can directly start the TS-InSAR process. The entire DefoEye processing workflow is executed within a working directory, where all necessary processing files and output data are stored. The workflow begins by setting up a top-level directory structure in the working directory and uses symbolic linking to reference S1 data files from the base directory, thereby significantly reducing disk space usage and processing time. DefoEye also automatically downloads the required SRTM DEM based on the spatial extent of the S1 data, eliminating the need for users to manually enter boundary coordinates.

### *2.2.2. Co-registration*

This step begins with generating a baseline table for all images. Then, all secondary images are aligned to a selected reference image to create a geometrically aligned stack of SLC data. The reference image can be specified by the user. The software automatically recommends the best five images with median temporal and spatial baselines relative to the rest of the dataset, minimizing spatial and temporal decorrelation. This is important since alignment precision depends on coherence between the reference and secondary images (Li et al., 2022). A robust geometric registration method is used, integrating the golden search algorithm and polynomial refinement techniques to achieve sub-pixel alignment accuracy (Xu et al., 2017). DefoEye is designed to run multiple jobs in parallel, enabling the simultaneous alignment of several secondary images. The number of parallel jobs for image alignment was set to 8 for all study areas in this study. The optimal number of parallel jobs in all the steps depends on the user's system hardware (mostly number of CPU threads). It is recommended to use fewer than all the available CPU threads. This process is carried out separately for each S1 sub-swath. After processing, the software performs a self-check by verifying the file sizes to ensure the workflow is completed successfully. The process is repeated for the images which are not processed successfully.

#### *2.2.3. Interferogram Formation*

This step constructs an interferogram network by connecting image pairs based on temporal and spatial baseline thresholds, which were set to 48 days and 250 meters in this study, to generate a network with eight connections for each image, as recommended by (Karami et al., 2025). Interferograms are generated through conjugate multiplication of image pairs. The phase of each interferogram contains not only the deformation signal but also other components that must be removed (Li et al., 2022). Using POD data and the SRTM DEM, the interferograms are flattened by eliminating phase contributions from Earth curvature, topography, and orbital errors. These steps are performed on full-resolution SLC images, which are often noisy and storage-intensive. To improve phase quality and reduce data volume, a two-stage filtering process is applied: first, a 2D Gaussian spatial filter with a 200 m wavelength is used, followed by a multi-looking process in both range and azimuth directions, set to 20 and 4, respectively in this study, which yields an output resolution of approximately 100 m, as recommended by (Khan and Xie, 2025). Finally, a Goldstein frequency-domain filter is applied as a secondary noise reduction step. All interferograms are computed independently for each sub-swath using consistent parameters. DefoEye also performs a self-check by verifying the number and size of the generated files to ensure successful completion of the process. The process is repeated for the interferograms which are not processed successfully. Parallel job processing is also provided at this stage, with number of parallel jobs set to 10 in this study for this stage.

#### *2.2.4. Unwrapping*

Interferogram phase values are confined within a modulo $2\pi$ range (between $-\pi$ and $+\pi$) (Ashraf et al., 2024). This wrapped phase is unwrapped using a 2D statistical cost network flow algorithm known as SNAPHU (Xu et al., 2017). Unwrapping is performed in regions with coherence higher than a threshold (set to 0.1 in this study) to avoid generating displacement maps with large missing information. InSAR measurements over water can compromise the unwrapping process due to the lack of coherence (Lazecký et al., 2020, Hrysiewicz et al., 2023). Therefore, a water mask, derived from the Global Self-consistent, Hierarchical, High-resolution Geography (GSHHG) database (Wessel and Smith, 1996), last updated in 2017, is applied prior to unwrapping. Users may also define an Area of Interest (AOI) by specifying a bounding box, limiting unwrapping and subsequent processing to the selected region. This reduces processing time and file size, especially when the AOI does not cover the full S1 sub-swaths. The unwrapping process is executed in parallel job processing mode (set to 10 in this study), allowing multiple interferograms to be unwrapped simultaneously.

Some unwrapped interferograms may contain errors due to high noise levels in the wrapped phase, often resulting from spatial and temporal decorrelation (Li et al., 2022). DefoEye automatically evaluates interferogram quality using triplet loop closure technique after unwrapping (Biggs et al., 2007, Morishita et al., 2020). Given three unwrapped interferograms ($\varphi_{12}$, $\varphi_{23}$, and $\varphi_{13}$), the loop phase ($\varphi_{123}$) is calculated as shown in Eq. (1). If all three interferograms are error-free, the loop phase should be close to zero. Interferograms that consistently produce a Root Mean Square (RMS) higher than a threshold (set to 1.5 radians in this study) in all associated loops are excluded from subsequent SBAS processing (Ghorbani et al., 2022, Morishita et al., 2020).

$$\varphi_{123} = \varphi_{12} + \varphi_{23} - \varphi_{13} \tag{1}$$

*2.2.5. Anchoring*

TS-InSAR displacement values are inherently relative in both time and space. In DefoEye, displacements are referenced temporally to the first image. Spatially, anchoring is also required to establish a zero-reference point for meaningful interpretation of deformation (Johnston et al., 2021). Spatial anchoring is not supported in the standard GMTSAR workflow. Based on the literature, two primary anchoring methods are supported: referencing to (1) a specific point or region or (2) the average phase (Haghshenas Haghighi and Motagh, 2024a, Xu et al., 2017). Users can manually define an arbitrary reference point by inputting latitude and longitude (e.g., a GNSS station with zero or minimum recorded deformation). A buffer can also be applied around the selected point to define a reference region, reducing the impact of noise and isolated pixel artifacts. If the chosen reference is itself undergoing displacement, all scene values will be biased, potentially leading to misinterpretation. Hence, DefoEye also supports average-phase referencing, where the average unwrapped phase is subtracted from each interferogram (Haghshenas Haghighi and Motagh, 2024a). This method is also used in other TS-InSAR packages like StaMPS (Hooper et al., 2004).

*2.2.6. TS-InSAR*

To estimate displacement values, a stack of unwrapped interferograms is inverted using a least squares adjustment, which employs coherence-weighted Small Baseline Subset (SBAS) method (Schmidt and Bürgmann, 2003, Berardino et al., 2003). Unlike traditional approaches that discard low-coherence pixels, this algorithm incorporates interferometric coherence values as weights in solving the least squares problem (Tong and Schmidt, 2016, Wang et al., 2022). Additionally, a temporal smoothness constraint (set to 1 in this study) is applied in the SBAS retrieval process to suppress unrealistically large temporal variations. Notably, parametric modeling is used for atmospheric correction of unwrapped interferograms. This process first estimates and removes the topographic signal caused by variations in atmospheric stratification. It then applies common scene stacking to filter out atmospheric noise using an iterative approach, which requires the number of iterations (set to 3 in this study) parameter (Tymofyeyeva and Fialko, 2015). This stage outputs the LOS deformation maps (in millimeters) at each S1 acquisition date.

*2.2.7. Analysis*

The generated deformation maps are geocoded to the WGS1984 coordinate system and saved as .tif and .grd files for user convenience. At this stage, users can specify the desired spatial resolution for the output in meters (set to 100 m in this study). Additionally, DefoEye includes a validation module that uses daily GNSS observations in tenv3 format from NGL (see Section 2.1 for details). GNSS observations are converted to LOS displacement values based on SAR look angles at each GNSS station. These are compared with average InSAR-derived LOS values within a 100 m radius of each GNSS station to reduce geolocation discrepancies between the two datasets.

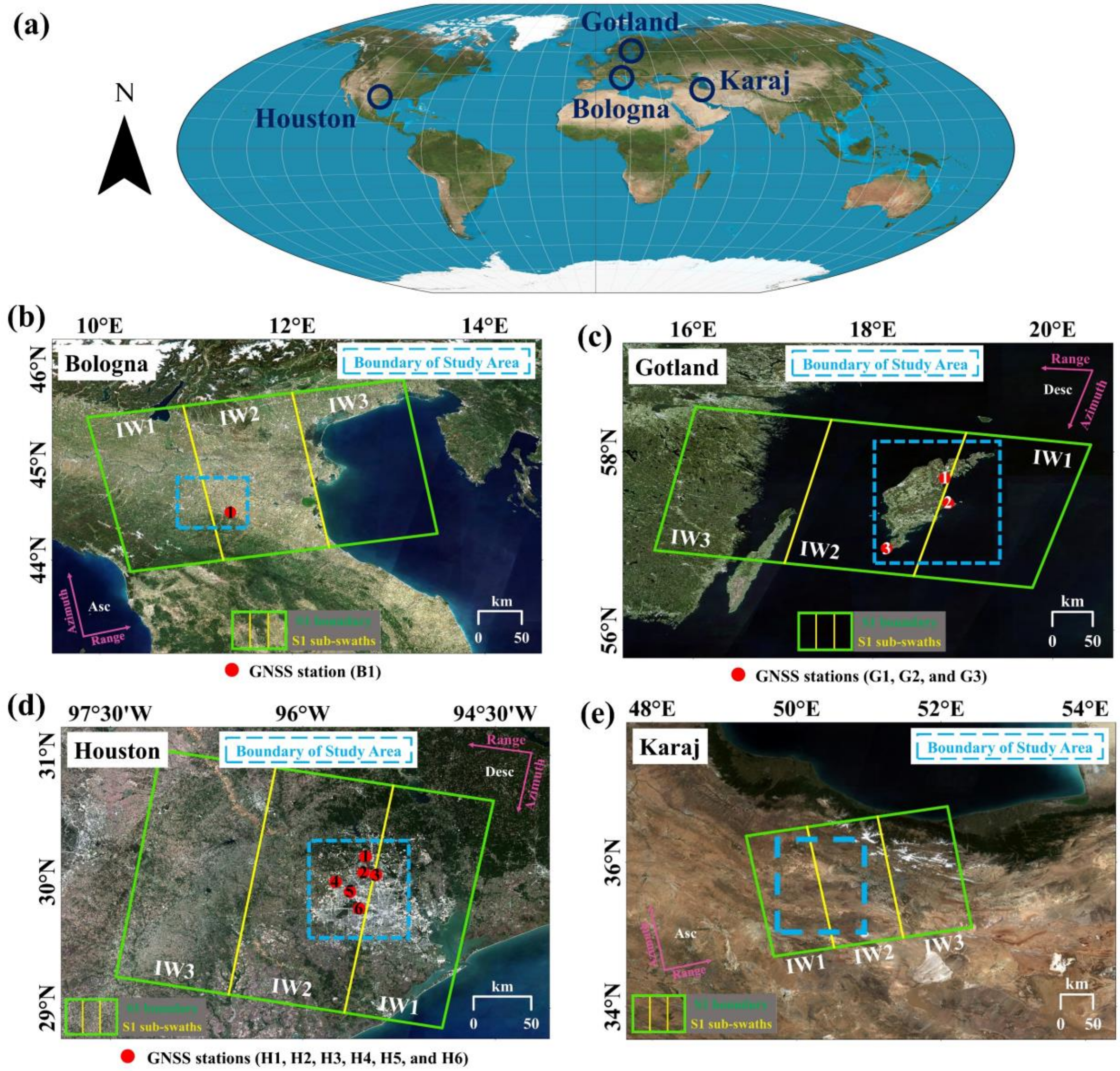

**Figure 2.** (a) Global locations of the study sites. (b), (c), (d), and (e) S1 scene boundaries, corresponding sub-swaths, and range/azimuth directions of the SAR images for Bologna, Gotland, Houston, and Karaj, respectively. The blue rectangles delineate the boundaries of the study regions. Red dots indicate the GNSS station locations (see Section 3.2) for each region (Bologna: B1, Gotland: G1–G3, and Houston: H1–H6).

# 3. Implementation and results

## 3.1. Study Areas

DefoEye (v1) performance in GD mapping was evaluated in four geographically distinct areas with differing geological settings, causes of GD, atmospheric and climatic conditions encompassing Bologna (Italy), Gotland (Sweden), Houston (USA), and Karaj (Iran), the global locations of which are illustrated in Figure 2a. These regions experience seasonal and long-term surface displacements caused by both natural and anthropogenic processes due to their environmental, geological, and anthropogenic settings. The first study area, Bologna, is in the Emilia-Romagna alluvial plain in northern Italy, covering the city of Bologna (Figs. 2b and 2e). The region is highly industrialized and densely populated, hosting one of the largest concentrations of oil and gas fields in Europe (Cenni et al., 2013). It also contains a complex multi-aquifer system, with groundwater sources ranging from freshwater to brackish and saltwater aquifers. The second study area, Gotland, is Sweden's largest island, located in the Baltic Sea (Figs. 2c and 2f). Its geology consists of lime-dominated regions and areas composed of marl, where interlayering of these materials is also common (Erlström et al., 2022). Large peatlands in the mid-northern and mid-southern parts of the island have been converted into farmland, leading to gradual GD as the dried peat oxidizes and compacts over time. Additionally, agricultural drainage systems accelerate groundwater loss by quickly diverting water into the ocean rather than allowing it to slowly recharge the groundwater table, significantly reducing groundwater recharge rates. Houston, the third study site, is the largest city in Harris County, with a population exceeding 2 million (Figs. 2d and 2g). It is among the largest metropolitan areas in the USA and is situated in the southeastern part of Texas, along the Texas Gulf Coast at the northwestern edge of the Gulf of Mexico. The Houston region has been experiencing GD since the 1920s. Studies suggest that subsurface fluid extraction, including oil, gas, and groundwater, is a primary contributor to GD in Houston and its surrounding areas (Khan et al., 2014). This is because the Harris region contains several productive oil and gas fields, further intensifying the GD. Additionally, groundwater serves as the primary water source for municipal, agricultural, and industrial use in the Houston area. The fourth study area (Karaj) is located in Iran, adjacent to the capital city, Tehran. In this region, excessive extraction of groundwater has caused severe GD (Haghshenas Haghighi and Motagh, 2024b). With the growing population and climate change–induced variations, the recharge of groundwater resources has also been hindered, further intensifying subsidence in the area. These diverse GD-causing processes, combined with the differing climate conditions and geological characteristics, make the four regions a suitable testbed for assessing DefoEye's performance.

## 3.2. S1 data and GNSS stations

At the time preparing S1 data for this research, S1A was the only operational satellite; therefore, only its data were used. Figures 2b-e illustrate the scene boundaries and sub-swaths covering each study region. Table 1 provides a summary of the S1A datasets used for TS-InSAR analysis across the study sites. The period for S1A data was selected from 2020 to 2024. For validation of DefoEye results, TS measurements from 10 GNSS stations were utilized: B1 in Bologna, G1–G3 in Gotland, and H1–H6 in Houston (Figs. 2b-d). The location of these stations is also provided in Table 1. These stations were selected based on their continuous observations during the study period (2020–2024) and their location within coherent, gap-free regions in the final GD maps, with a 100-meter radius applied to reduce the influence of isolated pixels. No GNSS stations were available in the Karaj region; therefore, the DefoEye results for this area were evaluated by comparison with outputs from other TS-InSAR processing platforms.

**Table 1.** Detailed S1 data and GNSS station locations for each study site. Asc denotes ascending, and Desc denotes descending acquisition modes of the S1 imaging trajectory. Either Asc or Desc mode was selected only when it ensured complete coverage of the study area (Fig. 2).

| | **Study site** | | | |
|---|---|---|---|---|
| | **Bologna** | **Gotland** | **Houston** | **Karaj** |
| **Area [× 1000 km$^2$]** | 1.2 | 11 | 10.2 | 11.5 |
| **Date of first S1 image** | 2020-01-03 | 2020-01-09 | 2020-01-05 | 2020-01-09 |
| **Date of last S1 image** | 2024-02-23 | 2024-02-29 | 2024-02-25 | 2024-02-29 |
| **Total Number of images** | 123 | 127 | 124 | 127 |
| **Imaging mode** | Asc | Desc | Desc | Asc |
| **Sub-swath Numbers** | 1, 2 | 1, 2 | 1, 2 | 1, 2 |

| S1 acquisition track | Path: 117, Frame: 143 | Path: 22, Frame: 400 | Path: 143, Frame: 492 | Path: 28, Frame: 112 |
|---|---|---|---|---|
| **Average incidence angle** | 36.4º | 36.1º | 36.5º | 39.1º |
| **GNSS stations (latitude, longitudes)** | B1 (44.5º, 11.35º) | G1 (57.70º, 18.79º) G2 (57.42º, 18.38º) G3 (56.92º, 18.15º) | H1 (30.19º, -95.51º) H2 (30.07º, -95.52º) H3 (30.05º, -95.43º) H4 (29.99º, -95.74º) H5 (29.91º, -95.63º) H6 (29.78º, -95.56º) | No GNSS data available. The results were compared to other TS-InSAR processing software. |

## 3.3. Other TS-InSAR processing software

For regions with available GNSS stations (Bologna, Gotland, and Houston), GNSS-based validation was the most reliable method for assessing DefoEye's results. However, since no GNSS data were available for Karaj, the DefoEye results were compared with those obtained from a widely used Python-based software, MintPy, for TS-InSAR analysis (Yunjun et al., 2019). MintPy requires a stack of unwrapped interferograms, which in this study were generated using Hybrid Pluggable Processing Pipeline (Hogenson et al., 2016). The same zero-reference points, processing parameters, and interferogram networks were used in the MintPy analysis as in the DefoEye processing to ensure a fair comparison between the two platforms. In MintPy, tropospheric correction was applied using ERA5 data via the Python-based Atmospheric Phase Screen estimation tool (PyAPS), and topographic residual correction was performed using a second-order polynomial temporal deformation model.

## 3.4. Results and validation

### *3.4.1. SBAS networks*

Figure 3 presents the interferogram networks and spatial baseline distributions for all four study sites. The interferogram networks in each region appear dense and well connected. The spatial baseline distributions reveal that most interferograms have short baselines (0–150 m), with fewer pairs at longer baselines, minimizing spatial decorrelation due to acquisition geometry. The software's recommendation was used for selection of the reference image, resulting in optimally chosen reference images (green dots) located near the median temporal and spatial baselines to enhance image alignment accuracy by reducing spatial and temporal decorrelation effects. Among the four sites, Karaj and Houston retain the highest number of interferograms (466 and 460, respectively) after loop-closure filtering, suggesting fewer unwrapping errors. This is likely due to these study areas having higher-coherence regions, with Karaj having less vegetation cover and Houston being predominantly urban, both of which improve phase unwrapping accuracy. In contrast, Bologna and Gotland had more interferograms removed (62 and 49, respectively), primarily due to lower coherence in areas influenced by vegetation, such as agricultural lands, which can degrade unwrapping quality.

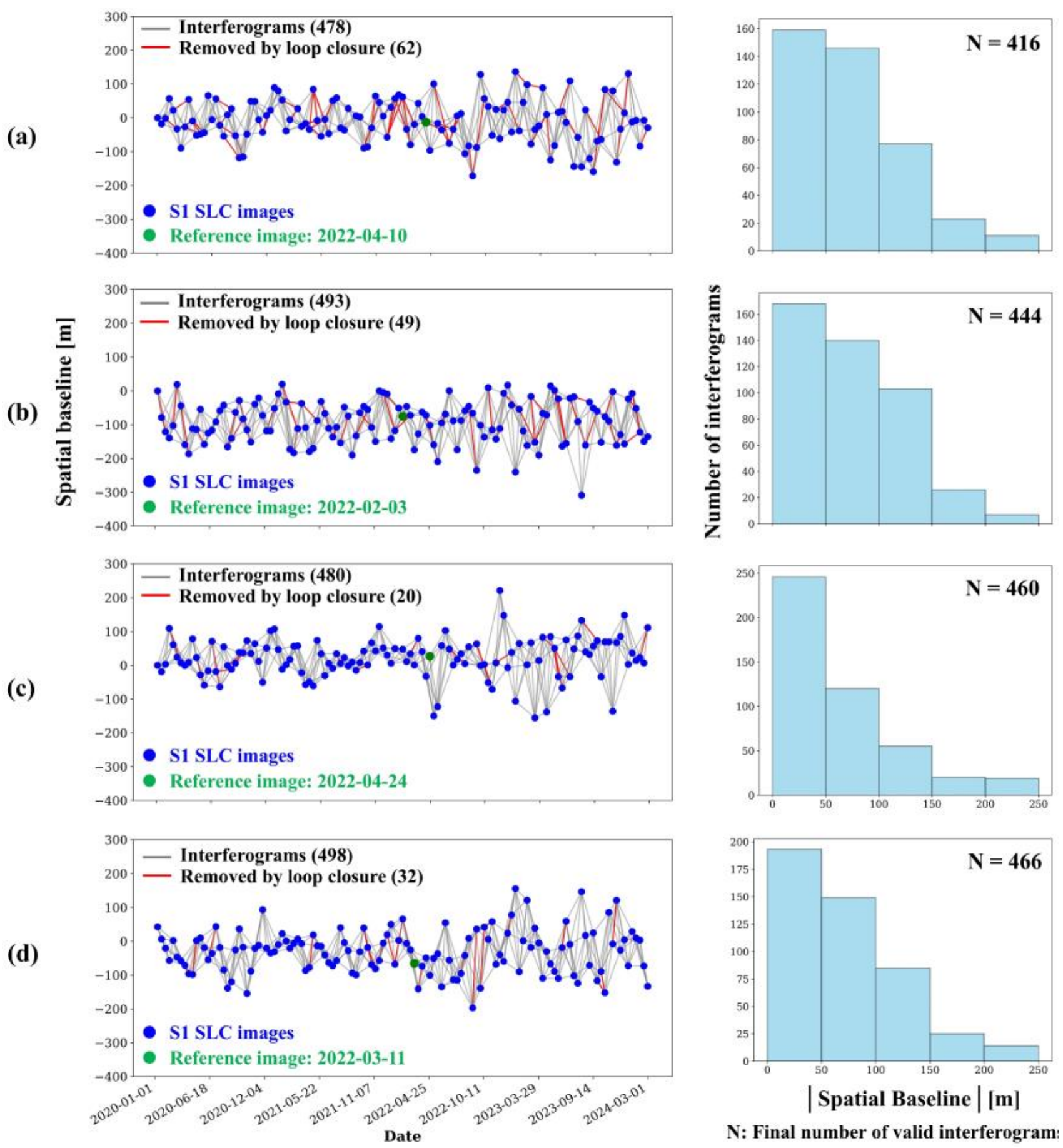

**Figure 3.** Interferogram networks (left-side plots) and spatial baseline distribution of interferograms (right-side histograms) for (a) Bologna, (b) Gotland, (c) Houston, and (d) Karaj.

*3.4.2. GNSS-based validation of DefoEye*

The displacement velocity maps for the three study sites with available GNSS data (Bologna, Houston, and Gotland) are presented in Figure 4. The results indicate that both positive (uplift) and negative (subsidence) LOS displacements were detected. In Bologna, subsidence is concentrated in the northwestern part of the city, consistent with findings from recent studies (Bitelli et al., 2025). In Houston, the velocity map reveals ongoing subsidence in the western and northern parts of the city, while the southeastern region shows negligible subsidence and even minor uplift (Liu et al., 2022). It is important to note that the density of measurement points varies depending on land cover and coherence. The maps show a high density of points in urban and industrial areas, where coherence is better maintained, whereas areas with dense vegetation exhibit sparse measurements due to decorrelation effects.

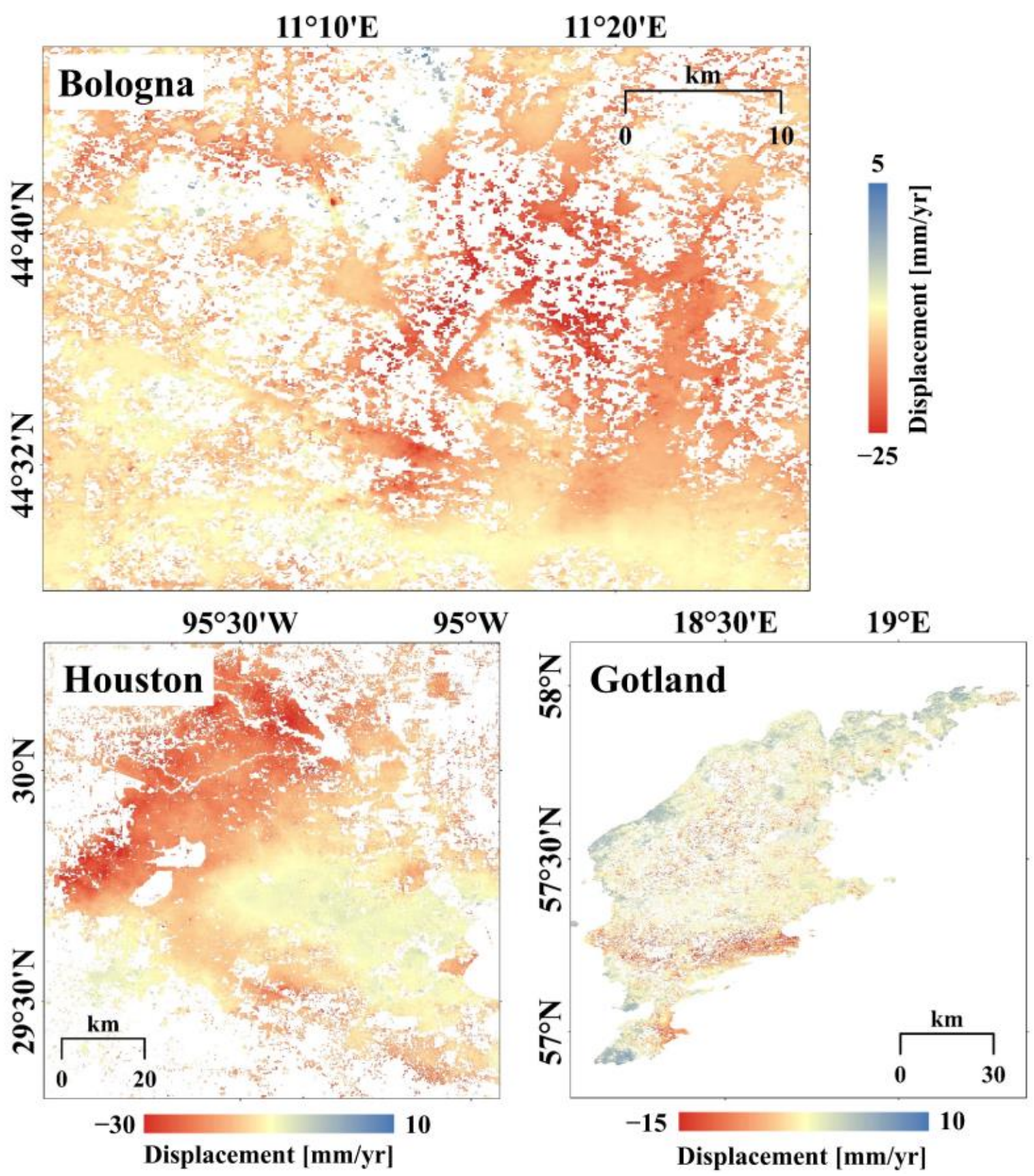

**Figure 4.** Displacement velocity maps for Bologna, Houston, and Gotland for GNSS-based assessment of DefoEye.

For quantitative validation, the Root Mean Squared Error (RMSE) and Pearson Correlation Coefficient (PCC) between GNSS and InSAR measurements were computed. Moreover, their histogram distributions were compared using the Kolmogorov-Smirnov (KS) statistical test, which provides two key outputs: the KS-statistic and the p-value (Fasano and Franceschini, 1987). A lower p-value (or higher KS statistics) below 0.01 indicates that the two distributions are significantly different at the 1% significance level, corresponding to a 99% confidence level.

The RMSE between InSAR and GNSS falls within a low range of 4.3 to 11.9 mm over the four-year period from 2020 to 2024. The slopes are also closely matched, with the smallest difference observed at H5, where the InSAR-derived slope is -4.64 mm/yr and the GNSS-derived slope is -4.56 mm/yr. Moreover, the TS-InSAR results generated by the proposed software closely aligned with GNSS displacement trends, capturing both the timing and direction of LOS displacement. The results indicate that whenever GNSS detects movement along the LOS, InSAR reflects a similar pattern with no noticeable time offsets; peaks and troughs in displacement occur nearly simultaneously in both datasets. This strong agreement is quantitatively supported by high PCC values, ranging from 0.63 to 0.95 across all stations. These results also indicate that the long-term displacement rates measured by InSAR are generally well-aligned with GNSS rates, confirming the reliability of DefoEye for long-term TS-analysis monitoring.

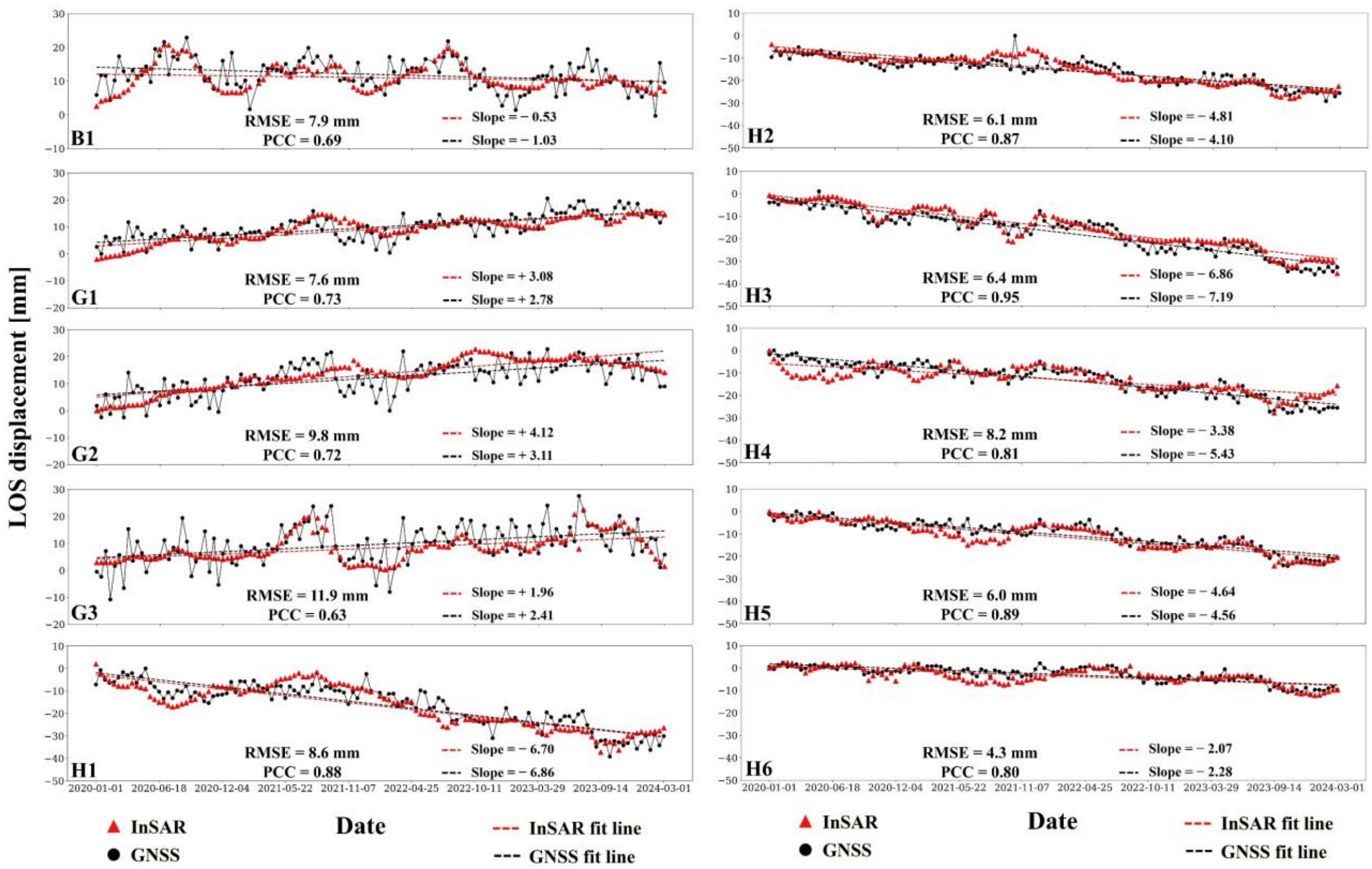

**Figure 5.** Comparison of InSAR and GNSS displacements in the LOS direction for 10 stations across three of the study areas with available GNSS data (Bologna: B1, Gotland: G1–G3, and Houston: H1–H6). The locations of these stations are shown in Figure 2 and are also listed in Table 1.

Figure 6 presents a statistical comparison of InSAR and GNSS LOS displacement distributions across 10 stations in the three study regions. The fitted Gaussian distributions on both InSAR and GNSS histograms are centered around similar values for most of stations, resulting in overlapping peaks, which indicate minimal bias between the two datasets. The smallest difference in average displacement (μ) is observed at H4, with a difference of 0.04 mm (InSAR: -12.75 mm, GNSS: -12.71 mm), while the largest difference is 2.06 mm at H3 (InSAR: -14.75 mm, GNSS: -16.81 mm). Additionally, the standard deviation (σ) values are closely matched across all stations, with a minimum difference of 0.14 mm at H6 and a maximum difference of 1.9 mm at G3. Furtheremore, most stations exhibit a low KS statistic and a relatively large p-value, indicating that the InSAR and GNSS displacement distributions are not statistically different. Stations B1, G1, G2, H1, H4, and H5 demonstrate a strong agreement between InSAR and GNSS measurements, while stations G3, H2, H3, and H6 exhibit marginal statistical similarities; however, none strictly meet the criteria for rejection (p-value ≤ 0.01). Overall, the findings from Figures 5 and 6 highlight the strong consistency between TS-InSAR results and GNSS observations, confirming the effectiveness of DefoEye for long-term GD monitoring.

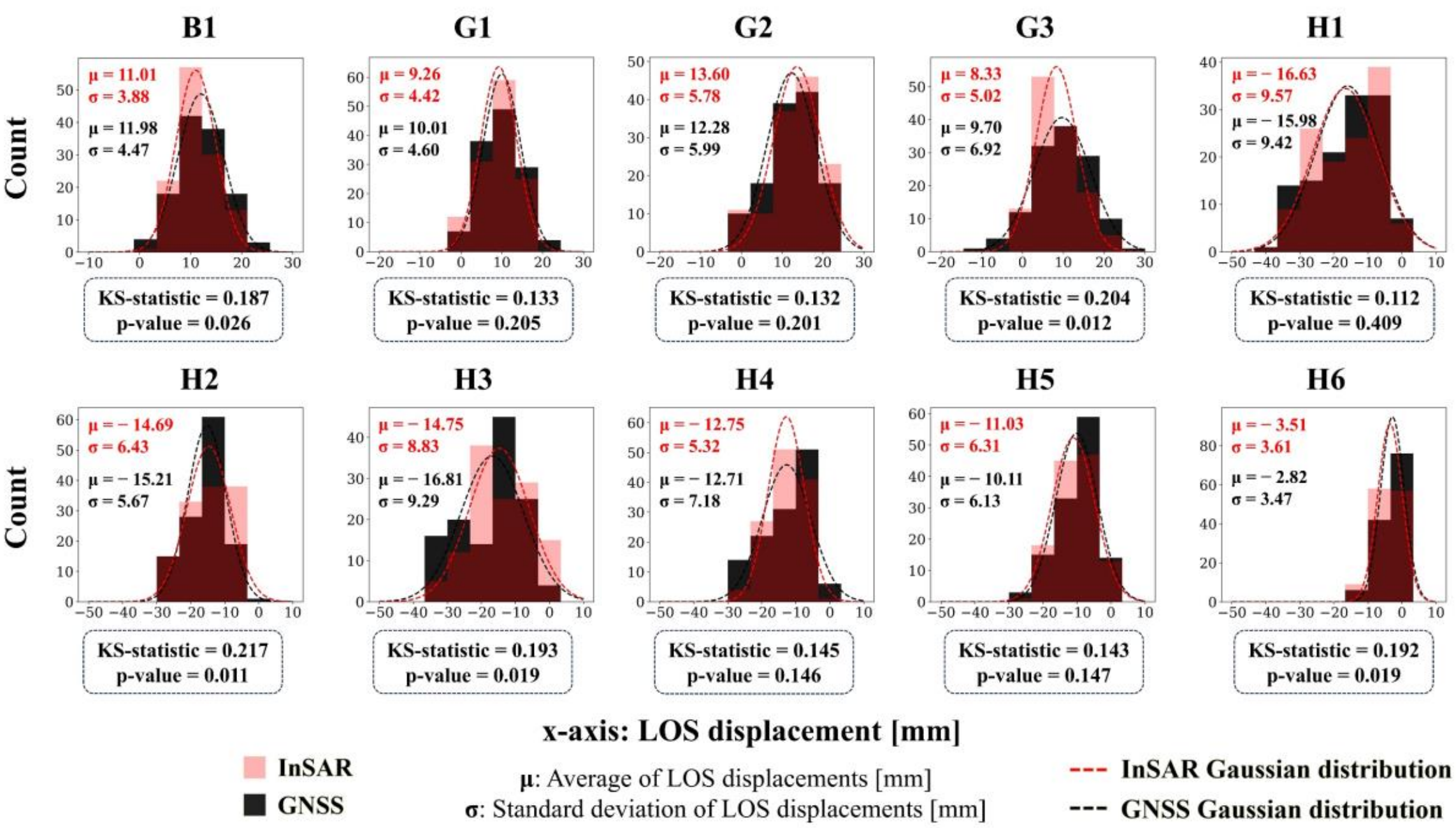

**Figure 6.** Histogram comparison of InSAR and GNSS LOS displacement distributions for 10 stations across the study areas (Bologna: B1, Gotland: G1- G3, and Houston: H1- H6). The Gaussian distribution of each dataset is overlaid (dashed lines), and the KS test results are also provided. The locations of these stations are shown in Figure 2 and listed in Table 1.

DefoEye integrates both interferogram network pruning and spatial anchoring into the TS-InSAR processing workflow. Figure 7 compares the LOS deformation rates [mm/yr] of average InSAR values within a 100-meter buffer around GNSS stations with GNSS-derived LOS deformation rates under three scenarios: (1) without pruning or anchoring, (2) with anchoring only, and (3) with both pruning and anchoring. The results show that the RMSE decreases to 0.80 mm/yr when both steps are applied, while the absence of both yields the highest RMSE of 2.45 mm/yr. A similar trend is observed in the $R^2$ values. These results underscore the importance of incorporating both interferogram network pruning and spatial anchoring in TS-InSAR analysis.

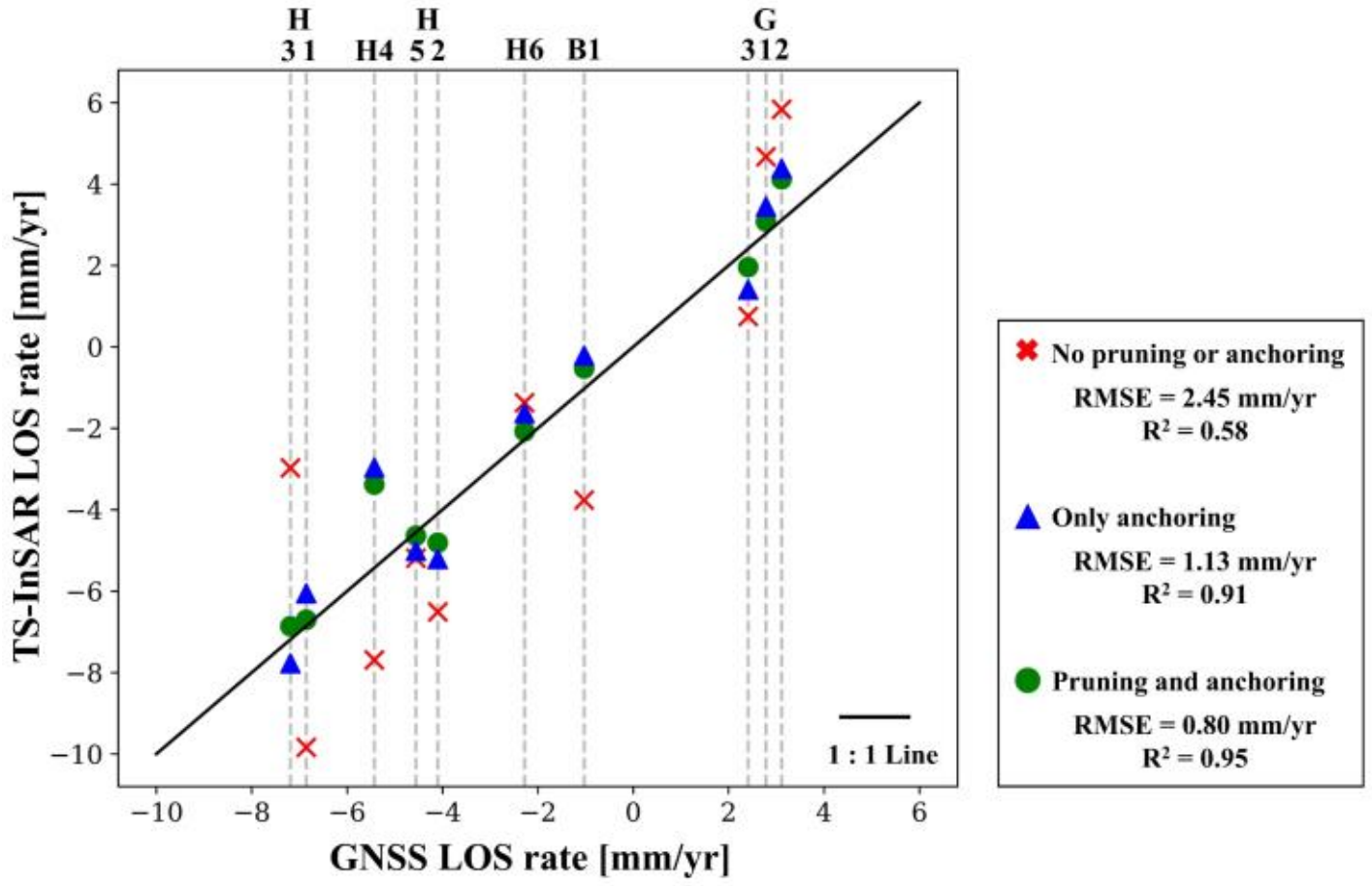

**Figure 7.** Comparison of LOS deformation rates [mm/yr] between average TS-InSAR rates within a 100-meter buffer around GNSS stations and GNSS-derived LOS rates under three scenarios: (1) without pruning or anchoring, (2) with anchoring only, and (3) with both pruning and anchoring applied. Gray dashed lines represent GNSS stations.

*3.4.3. Comparison with MintPy*

As mentioned in Section 3.3, the DefoEye results were compared with MintPy, as shown in Figure 8. The final displacement velocity maps are highly similar, as indicated by the high PCC of 0.98 and the low RMSE of 4.6 mm/yr. Furthermore, both processing platforms yield comparable distributions of displacement velocities, as shown by the scatterplots, where points lie predominantly along the 1:1 line. These results demonstrate that DefoEye produces outputs consistent with those from the MintPy platform.

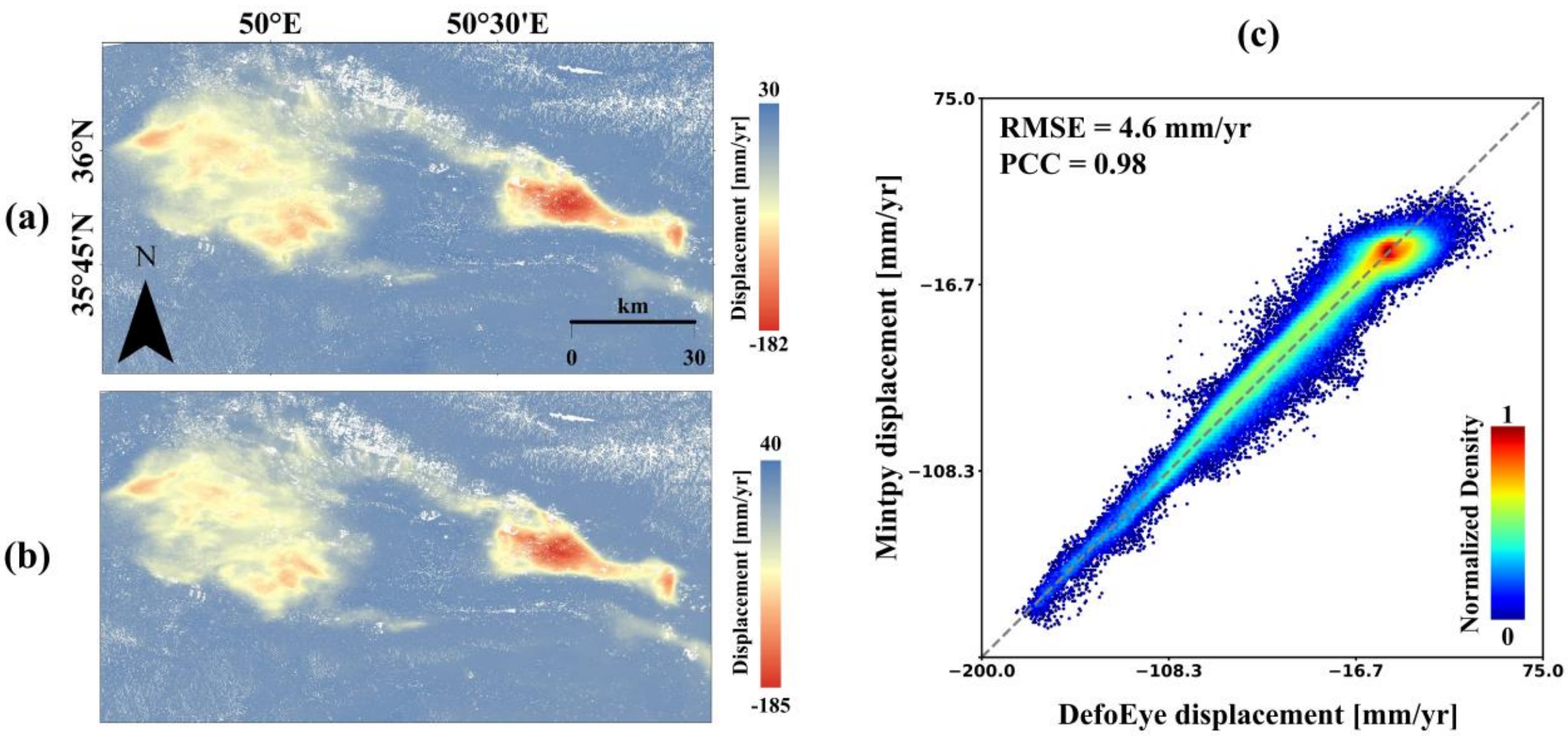

**Figure 8.** Comparison of LOS displacement velocity maps from (a) DefoEye and (b) MintPy. (c) shows the scatterplot of point-to-point displacement velocity values between the two platforms. The reference point in both (a) and (b) maps is (lat: 35.37, and lon:50.67). The gray dashed line in (c) represents the 1:1 correspondence line.

*3.4.4. Single vs parallel job execution*

Table 2 demonstrates the effectiveness of parallel job execution within DefoEye workflow. As shown, parallel processing significantly reduces computation time across all regions and processing steps. For instance, the image alignment step in Houston was reduced from 636 minutes to 196 minutes using workstation (a). Phase unwrapping showed the greatest variability in execution time among the processing steps. This is influenced by both regional coherence and area size, as AOI clipping is applied before this step. Bologna, the smallest region, had the shortest unwrapping time (36 minutes for workstation (a)) compared to Gotland and Houston with similar areas. However, Gotland required significantly more time compared to Houston since Houston region is dominated by man-made structures with higher coherence, which is more straightforward for unwrapping procedure. It is important to note that Table 2 focuses solely on steps where parallel job processing is supported. Additional time savings are achieved throughout the pipeline due to automated directory structuring, and seamless command execution within DefoEye.

**Table 2.** Execution times (in minutes) for different parallel processing-supported steps in DefoEye using single and parallel job processing across the three study areas. The reported times reflect the specific processing parameters described in Section 3; variations in parameter settings may lead to different execution times. All processes were performed on two different workstations: (a) one equipped with 64 GB of RAM and an Intel Core i9-13900 CPU, and (b) another with 8 GB of RAM and an Intel Core i7-6700 CPU.

| Study site | Workstation | Bologna | | Gotland | | Houston | | Karaj | |
|---|---|---|---|---|---|---|---|---|---|
| S1 data volume | | 327 GB | | 335 GB | | 344 GB | | 345 GB | |
| Processing type | | Single | Parallel | Single | Parallel | Single | Parallel | Single | Parallel |
| Step: Image alignment [minutes] | (a) | 619 | 193 | 634 | 195 | 636 | 196 | 652 | 199 |
| | (b) | 1627 | 549 | 1776 | 583 | 1884 | 604 | 1976 | 604 |
| Step: Interferogram Computation and filtering [minutes] | (a) | 837 | 163 | 838 | 168 | 842 | 176 | 845 | 177 |
| | (b) | 1759 | 483 | 1802 | 491 | 1806 | 498 | 1808 | 498 |
| Step: Merging of the interferograms [minutes] | (a) | 93 | 47 | 102 | 51 | 93 | 49 | 105 | 50 |
| | (b) | 254 | 63 | 306 | 84 | 266 | 89 | 308 | 88 |
| Step: Phase Unwrapping [minutes] | (a) | 223 | 36 | 537 | 129 | 436 | 75 | 449 | 78 |
| | (b) | 619 | 75 | 1198 | 276 | 952 | 204 | 1037 | 215 |
| Output volume | | 501 GB | | 538 GB | | 541 GB | | 543 GB | |

# 4. Discussion

## 4.1. Analysis of DefoEye

This paper introduced DefoEye, a software that wraps GMTSAR to streamline the TS-InSAR processing chain. DefoEye features an easy-to-use GUI, addressing a common limitation in TS-InSAR processing software, most of which lack a GUI and can therefore be difficult for users with diverse technical backgrounds. As illustrated in Figure 1, DefoEye incorporates several new features into the GMTSAR workflow. The entire process is structured sequentially, with two key stages added: interferogram phase pruning and anchoring of unwrapped interferograms. Results shown in Figure 7 highlight the significance of these steps, as the RMSE and PCC between InSAR-derived and GNSS-derived results improved after their implementation compared to the original GMTSAR workflow, which does not take them into account. Interferogram phase pruning removes noisy interferograms with high unwrapping errors, directly enhancing final accuracy (Morishita et al., 2020). Anchoring the unwrapped interferograms resolves the $2n\pi$ ambiguity in the unwrapped phase (where n is an integer), leading to more accurate final displacement estimates (Hogenson et al., 2016, Yunjun et al., 2019).

DefoEye employs parallel processing in several key steps, as shown in Figure 1. This parallel job execution significantly accelerates the processing time for various stages, as summarized in Table 2. For example, it sped up the image alignment stage by more than three times and interferogram formation and filtering by more than five times, in all regions and on both tested workstations with different processing powers. The analysis was conducted on two different workstations with varying processing capabilities, and the results showed that processing time was greatly reduced when parallel processing was used compared to single-threaded execution. In addition to these gains, the complete implementation of the GMTSAR workflow, from the general setup to the export of results, through structured and automated Python scripts further reduces processing time and minimizes user intervention.

DefoEye, a Python-based TS-InSAR platform, was compared with MintPy, which is also widely used in the InSAR community due to its Python-based design that enhances applicability and usability. However, MintPy does not support a complete end-to-end TS-InSAR workflow; instead, it requires a pre-prepared stack of unwrapped interferograms for analysis. So, it is relied on external software. The need to rely on additional software not only complicates the process for users but also increases processing time. This limitation does not allow long-term TS-InSAR processing, and with those credits, the users can only process a small number of scenes, making it infeasible for large-scale studies. Moreover, incompatibilities in data formats between different software packages and MintPy can lead to technical issues. In contrast, DefoEye offers a fully integrated end-to-end processing workflow—also Python-based like MintPy—along with the added advantage of a GUI, which MintPy lacks. The comparison results (Figure 8) show that both DefoEye and MintPy workflow produced very similar deformation estimates. This demonstrates that DefoEye provides a reliable unified platform for TS-InSAR analysis, whose performance is validated not only against GNSS data but also by its close agreement with results from widely used software such as MintPy.

### 4.2. Comparison to similar studies

For further validation, GNSS-based validation of DefoEye is compared against other studies that utilized GNSS measurements for validation (Table 3). The number of GNSS stations used for validation varies across studies, with some incorporating only a single station (Naghibi et al., 2022, Radman et al., 2021, Ghorbani et al., 2022), while others utilize a more extensive GNSS network (e.g., (Neely et al., 2019), with 45 stations. RMSE values reported in previous studies range from 1 mm to 12.8 mm, and the RMSE obtained using the proposed software falls within this range (4.3 mm to 11.9 mm). This proves that DefoEye can be reliably used for long-term TS-InSAR analysis using S1 data.

**Table 3.** Summary of studies evaluating InSAR-derived displacements using GNSS data. All studies utilized S1 data for TS-InSAR analysis. While some of the reviewed studies may have incorporated additional GNSS points, the table lists only the GNSS stations directly used for validation. For further details on each study, refer to the corresponding reference.

| References | Temporal period | Number of GNSS stations | Reported RMSE |
|---|---|---|---|
| Neely et al. (2019) | 2014-2017 | 45 | 3.3 to 11.4 |
| Yalvac (2020) | 2014-2018 | 2 | 1 to 5 |
| Radman et al. (2021) | 2014-2019 | 1 | 2.2 |
| Ghorbani et al. (2022) | 2017-2019 | 1 | 6.76 |
| Naghibi et al. (2022) | 2014-2019 | 1 | 12.8 |
| Tang et al. (2023) | 2017- 2020 | 7 | 8 |
| Su et al. (2023) | 2014- 2022 | 21 | 10.6 |
| DefoEye | 2020- 2024 | 10 | 4.3 to 11.9 |

## 5. Conclusion

With the increasing availability of freely accessible SAR data, there is a growing demand for streamlined TS-InSAR analysis. To this end, this paper introduced DefoEye, a Python-based software that wraps GMTSAR and offers facilitated TS-InSAR analysis of S1 data with parallel job execution across multiple steps. It also incorporates interferogram network pruning and flexible anchoring methods to improve the final estimation accuracy of surface displacements. The GNSS-based performance analysis of DefoEye across three study sites—Bologna (Italy), Gotland (Sweden), and Houston (USA)—using 10 GNSS stations confirmed its accuracy and reliability. Moreover, for Karaj (Iran), where no GNSS data were available, the results showed that both approaches produced similar outcomes, with DefoEye achieving this through a unified processing workflow without the need for external software. DefoEye is publicly available, with unrestricted access enabling broad applicability to various fields such as landslide monitoring, land subsidence assessment, infrastructure stability analysis, groundwater studies, mining operations evaluation, and the development of early warning systems. By offering a globally applicable, and end-to-end TS-InSAR processing solution, this paper demonstrates that DefoEye can deliver reliable results to the InSAR research community, facilitating the identification of potential hazards, enhancement of regional safety, and implementation of effective mitigation strategies worldwide.

### Disclosure statement

The authors declare that they have no relevant financial or non-financial interests to disclose that could have influenced this work.

### Funding information

This work was supported by the Swedish Research Council for Sustainable Development (Formas) under grant no. 2022-02148.

### Acknowledgments

The authors express their gratitude to ESA and ASF for providing free access to S1 data, as well as to NASA for making the SRTM DEM available. We also extend our thanks to NGL for providing GNSS data. Special thanks are extended to the GMTSAR development team for developing and maintaining this open-source InSAR processing software.

### Authors' contributions

**Alireza Taheri Dehkordi:** Conceptualization, Methodology, Programming and Software Development, and Writing – Original Draft Preparation. **Hossein Hashemi and Amir Naghibi:** Supervision, Conceptualization, and Writing – Review and Editing. All authors have read and approved the final version of the manuscript for publication.

### Software and Data availability statement

All codes and data supporting the findings of this study are freely available under a CC-BY license. The DefoEye software is also openly available at https://github.com/ATDehkordi/DefoEye, with a detailed step-by-step tutorial. Sentinel-1 SAR data were obtained from the Alaska Satellite Facility SAR Distributed Active Archive Center (ASF DAAC) (https://asf.alaska.edu/asfsardaac/, last accessed March 2025).